\documentclass[12pt]{article}
\usepackage[]{graphicx}
\usepackage{caption}
\usepackage{subcaption}
\usepackage{cite}
\usepackage{amssymb}
\usepackage{amsmath}

\title{Properties of a new quasi-axisymmetric configuration}
\author{S. A. Henneberg, M. Drevlak, C. N\"{u}hrenberg,\\
	 C. D. Beidler, 
	Y. Turkin, J. Loizu, P. Helander\\
	\small{Max-Planck-Institut f\"{u}r Plasmaphysik, Wendelsteinstr. 1, 17489 Greifswald	}\\
 \small{Sophia.Henneberg@ipp.mpg.de}}
\vspace{10pt}
\date{\today}  
\begin{document}

\maketitle
\begin{abstract}
A novel, compact, quasi-axisymmetric configuration is presented which exhibits low fast-particle losses and is stable to ideal MHD instabilities. The design has fast-particle loss rates below 8\% for flux surfaces within the half-radius, and is shown to have an MHD-stability limit of a normalised pressure of $\langle\beta\rangle=3\%$ where $\langle\beta\rangle$ is volume averaged. The flux surfaces at various plasma betas and currents as calculated using the SPEC equilibrium code~\cite{SPEC_3D} are presented. Neoclassical transport coefficients are shown to be similar to an equivalent tokamak, with a distinct banana regime at half-radius. An initial coil design study is presented to assess the feasibility of this configuration as a fusion-relevant experiment.
\end{abstract}
\section{Introduction}
Quasi-axisymmetric (QA) equilibria possess, similarly to tokamaks, a magnetic field strength that is symmetric in the toroidal Boozer coordinate, and thus share many neoclassical properties with tokamaks. These configurations are capable of being compact thanks to their relatively large bootstrap current which provides a source of rotational transform, in addition to that from the coils. On the other hand, the similarity to stellarators provides potential benefits: QA configurations can run in steady state potentially without any current drive and there is evidence from other types of stellarators that disruptions can be avoided if the vacuum rotational transform, created solely by the coils, is sufficiently large~\cite{disruption1,W7AS,CTH1}. 
The related concept of quasi-helical symmetry was numerically proven in 1988~\cite{HrN88} and experimentally confirmed in a series of experiments on the HSX stellarator~\cite{HSX07,HSX07_PoP}.\\
Good confinement in quasi-symmetric fields is ensured by the existence of a third constant of the guiding-centre motion besides the energy and the magnetic moment to confine the guiding-centre orbits~\cite{Boozer95}. The guiding-centre Lagrangian in Boozer coordinates ($\psi$, $\theta$, $\phi$) ~\cite{Boozer81} only depends on the magnetic field strength rather than the direction of the magnetic field:
\begin{eqnarray}
	L = \frac{m}{2B^2} (I \dot{\theta} + G \dot{\phi})^2 + Z e (\psi \dot{\theta}-\chi \dot{\phi}) - \mu B - Z e \Phi
\end{eqnarray}
where $m$ is the mass, $B$ the magnetic field strength, $I$ and $G$ the toroidal and poloidal current, respectively, $Ze$ the charge, $\chi$ the poloidal flux, $\mu$ the magnetic moment, and $\Phi$ the electrostatic potential~\cite{Helander14}.\\
One obtains a third invariant (in addition to $\mu$ and the energy) if the magnetic field strength $B$ and the scalar potential $\Phi$ only depend on two of the magnetic coordinates (e.g. $\theta$ and $\psi$). The canonical momentum corresponding to the third magnetic coordinate ($\phi$) is then conserved:

\begin{eqnarray}
	\dot{p}_\phi= \frac{\partial L}{\partial \phi} =0
\end{eqnarray}
The first quasi-axisymmetric equilibria were presented in 1994~\cite{HrN94} and 1996~\cite{Garabedian96}. They were followed by many more designs such as NCSX~\cite{NCSX}, CHS-qa~\cite{CHSqa}, and ESTELL~\cite{ESTELL}, none of which have been constructed. 
The aim of the present design is slightly different.
Here we aim for a compact (aspect ratio of 3 to 4), MHD-stable configuration with small fast-particle loss rates, so that the device could, at least in principle, scale up to a reactor. A compact, MHD-stable equilibrium has been designed before (e.g. NCSX, CHS-qa), but we simultaneously require  improved fast-particle loss rates compared to previous designs.\\

The paper is organized as follows. In Sect.~\ref{methods} we describe the methods used to obtain the new configuration, including a description of the optimization code ROSE and the criteria which were targeted. The main section of this paper (Sect.~\ref{quasdex}) provides an overview of the new configuration including the geometrical properties of the new equilibrium design (Sect.~\ref{config}),  results of fast-particle loss-fraction calculations (Sect.~\ref{fast}), stability calculations (Sect.~\ref{stab}), preliminary results of islands and chaotic region development with varying plasma beta and current (Sect.~\ref{spec}), neoclassical transport and bootstrap calculations (Sect.~\ref{neo}), and a preliminary coil set (Sect.~\ref{coils}). The paper finishes with a summary and ideas for future work. 
\section{Optimization Methods}
\label{methods}
\begin{figure}[h]
	\centering
	\includegraphics[width=0.6\linewidth]{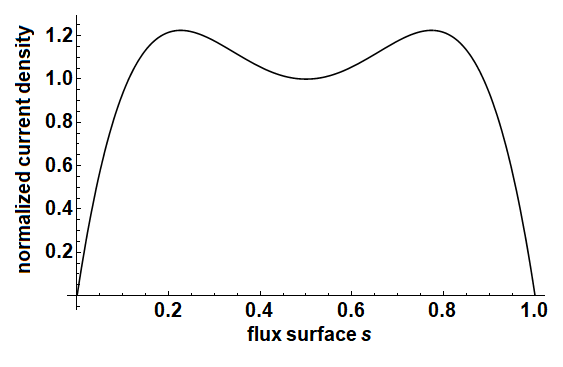}
	\caption{Normalized current density profile versus normalized flux $s$ for a bootstrap-like scenario. This profile is normalized to the total toroidal current; at the reactor size, the total toroidal current is approximately $2.5\,$MA.}
	\label{fig:curdens}
\end{figure}

\begin{figure}[h]
	\begin{subfigure}[t]{.45\textwidth}
		\centering
		\includegraphics[width=1.2\linewidth]{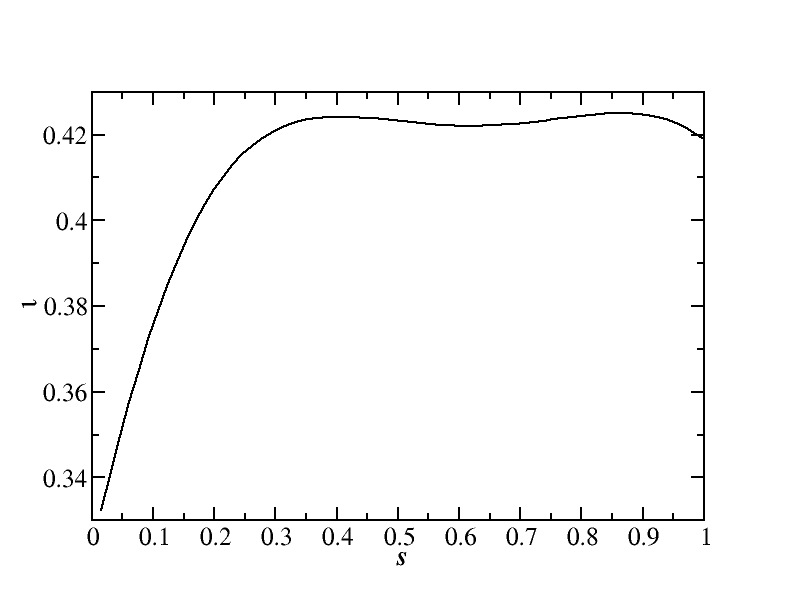}
		\caption{Rotational transform, $\iota$, versus normalized flux, $s$, with $\beta=3.5$\%.}
		\label{fig:iota}
	\end{subfigure}
	\hfill
	\centering
	\begin{subfigure}[t]{.45\textwidth}
		\centering
		\includegraphics[width=1.2\linewidth]{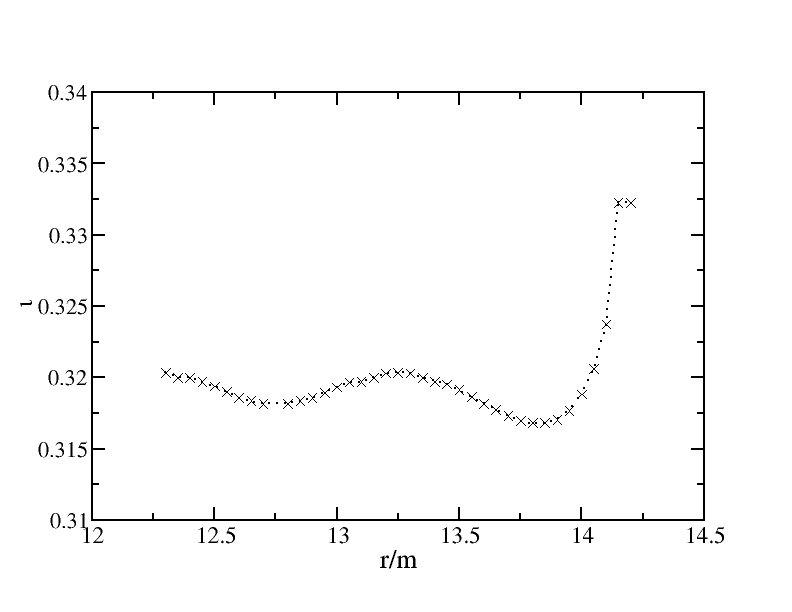}
		\caption{The vacuum rotational transform estimated with a current-carrying surface rather than coils.}
		\label{fig:vacIota}
	\end{subfigure}
	\caption{Rotational transform profiles for the new configuration at $\beta=3.5\%$ and $\beta=0\%$}
\end{figure}

\begin{figure}[h]
	\centering
	\includegraphics[width=0.8\linewidth]{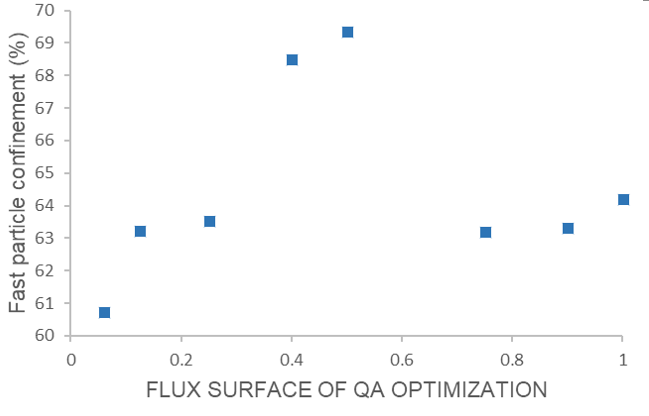}
	\caption{The effect of varying the location for the QA error optimization: The cumulative percentage of particles which remain inside the plasma for each optimized configuration versus location of QA optimization.}
	\label{fig:scan}
\end{figure}

The optimization tool ROSE (ROSE Optimizes Stellarator Equilibria~\cite{ROSE}) was exploited to examine the configuration space of ideal-MHD plasma boundaries. The standard optimization method is Brent's algorithm, which combines several root-finding methods such as the secant method, bisection method and a quadratic-inverse algorithm~\cite{Brent}. The cost function $f$ which is optimized is evaluated with the weighted sum method:
\begin{equation}
	f=\sum_i w_i (F_i-\tilde{F}_{i})^2 \, ,	\label{eq:weight}
\end{equation}
where $w_i$ are weights which have to be adapted for obtaining various configurations on the Pareto frontier\footnote{If the Pareto frontier is non-convex, there are points on it that cannot be found by this method.}. The latter is the set of optima where an optimum configuration is defined such that one cannot improve any criterion without worsening at least one other criterion. In Eq.~\ref{eq:weight}, $F_i$ is the value for the criterion, $i$, and $\tilde{F}_i$ is the corresponding target value. ROSE uses several other codes, including VMEC~\cite{VMEC}, a modified NESCOIL~\cite{NESCOIL}, and VM2MAG~\cite{VM2MAG}, and is capable of analysing many different criteria, including physical, geometrical and coil design properties.\\
For the plasma boundary optimization, we used the fixed-boundary equilibrium mode of VMEC. In this mode, VMEC represents the plasma boundary with a set of two Fourier series:
\begin{equation}
	r(u,v) = \sum r_{m,n} \cos(2\pi(m u-n N v)), \, z(u,v) = \sum z_{m,n} \cos(2\pi(m u-n N v)),
\end{equation}
where $r$ and $z$ are cylindrical coordinates, $u$ the VMEC poloidal and $v$ the VMEC toroidal angle.\\
For the configuration presented here, the chosen input parameters are:
\begin{itemize}
	\item aspect ratio $A=\frac{R}{a}$ where $R=r_{0,0}$ and $a=\sqrt{r_{1,0} z_{1,0}}$ are approximately the major and minor radii of the plasma, respectively,
	\item  number of field periods,
	\item volume-averaged plasma beta, $\beta=\frac{2\mu_0 p}{B^2}$, where $p$ is the plasma pressure and $\mu_0$ is the vacuum permeability.
\end{itemize}	
The targeted parameters for the optimization are: 
\begin{itemize}
	\item  rotational transform, $\iota$, at the magnetic axis and at the plasma boundary,
	\item vacuum rotational transform at the magnetic axis,
	\item  vacuum magnetic well $\frac{\partial}{\partial \psi} \int_{-\infty}^{\infty} \frac{\mathrm{d}l}{B} < 0$, where the integration is along a magnetic field line,
	\item the integrated absolute value of the Gaussian curvature of the plasma boundary, 
	\item and the quasi-axisymmetric error $\frac{\sqrt{\sum_{n\neq0,m} B_{m,n}^2}}{B_{00}}$ where the magnetic field strength is given by $B=\sum_{m,n} B_{m,n} \cos(m\theta+nN\phi)$.
\end{itemize}
We anticipate that the bootstrap current is sufficient to be the only net current in the plasma, e.g., no ohmic current is required. This is supported by preliminary calculations with the 1-D transport code NTSS~\cite{NTSS2006}. We therefore optimized with a bootstrap-like current density profile, see Fig.\ref{fig:curdens}. 
\\
We aimed for an external rotational transform above 0.3 and forced it to stay below 0.5 everywhere in the plasma, see Fig.~\ref{fig:iota}, to avoid the $\iota=1/2$ rational surface in the interest of MHD stability. \\
It can be shown that quasi-axisymmetry can only be achieved exactly on one single flux-surface~\cite{Garren,Plunk2018}, and we therefore minimized the quasi-axisymmetric error only on one designated flux-surface. Parameter scans are obtained by varying this flux-surface between the magnetic axis and the plasma boundary, see Fig.~\ref{fig:scan}. 
 The best fast particle confinement is achieved where the quasi-axisymmetric error is minimised at the flux surfaces $s=0.4$ and $s=0.5$. The normalized toroidal flux surface $s$ is defined as $s=\frac{\psi}{\psi_a}$ with $\psi_a$ the magnetic flux at the plasma boundary. Of those two configurations we chose the configuration with a higher vacuum magnetic well for stability reasons. We will report on these scans in greater depth in a future paper. The following section describes the final configuration, as illustrated in Fig.~\ref{contourstot}.
\section{The new design - an overview}
\label{quasdex}
\subsection{Configuration specification}
\label{config}
\begin{figure}
	\centering
	\begin{subfigure}[t]{.45\textwidth}
	\centering
	\includegraphics[width=1.0\linewidth]{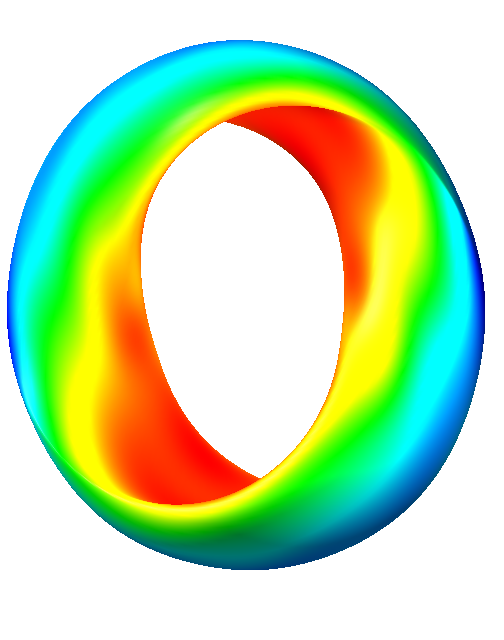}
	\caption{}
	\label{fig:threeD}
	\end{subfigure}
	\hfil
	\begin{subfigure}[t]{.45\textwidth}
	\centering
	\includegraphics[width=1.0\linewidth]{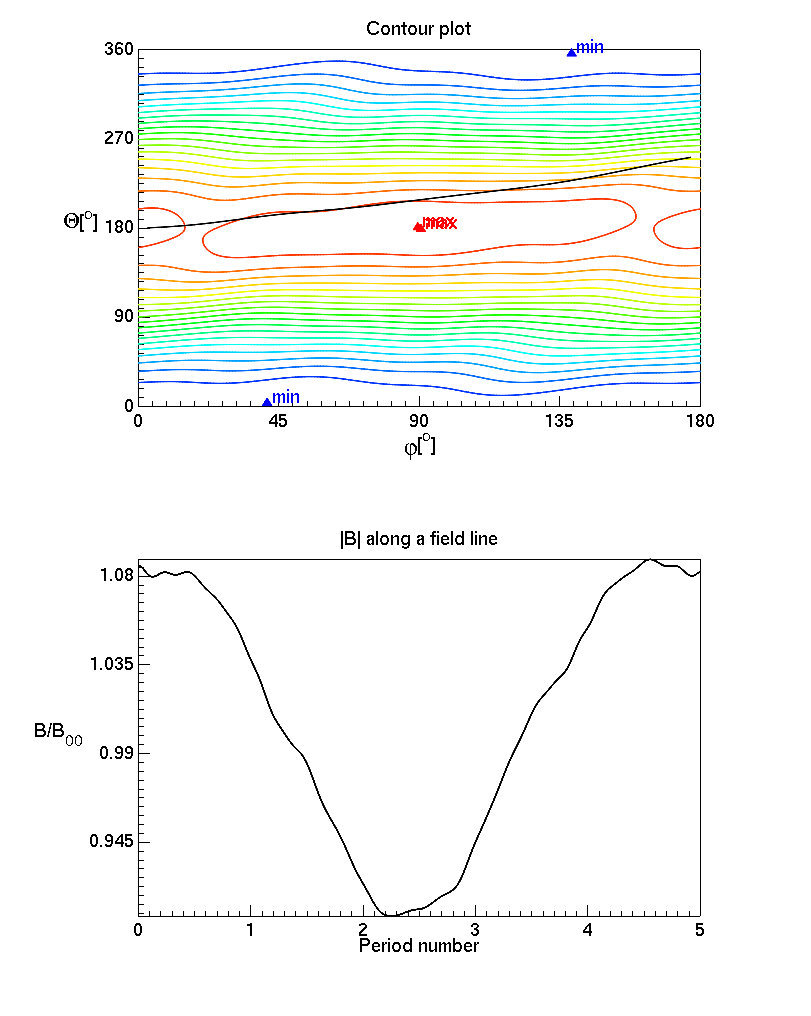}
	\caption{}
	\label{fig:contours}
\end{subfigure}
\caption{Magnetic field strength structure at the plasma boundary and at half radius. (a) The magnetic field strength on the plasma boundary of the new equilibrium design. (b) Top: The contours of the magnetic field strength at half radius ($s=0.25$) and a field line starting at $\varphi=0^\circ$ and $\theta=180^\circ$.  Bottom: Magnetic field strength along a magnetic field line starting at $\varphi=0^\circ$ and $\theta=180^\circ$. Each period is equivalent to a step size of $\Delta\varphi=180^\circ$. }
\label{contourstot}
\end{figure}

\begin{figure}[h]
	\centering
	\begin{subfigure}[t]{.3\textwidth}
		\centering
		\includegraphics[width=1.0\linewidth]{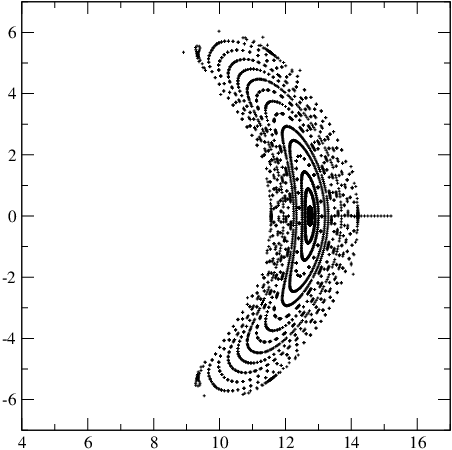}
		\caption{$\varphi =0$}
	\end{subfigure}
	\hfil
	\begin{subfigure}[t]{.3\textwidth}
		\centering
		\includegraphics[width=1.0\linewidth]{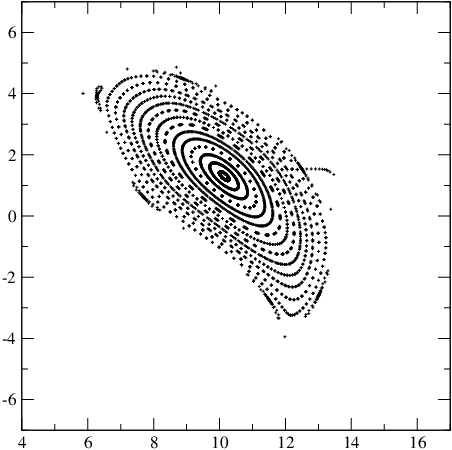}
		\caption{$\varphi = 45^\circ$}
		\label{fig:cutmid}
	\end{subfigure}
	\hfil
	\begin{subfigure}[t]{.3\textwidth}
		\centering
		\includegraphics[width=1.0\linewidth]{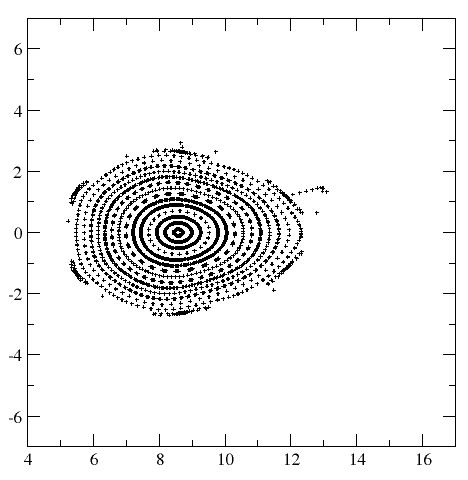}
		\caption{$\varphi=90^\circ$}
	\end{subfigure}
	\caption{Poincar\'{e} plots of the cross sections of the estimated vacuum field. These plots indicate nested magnetic surfaces. In vacuum, islands can potentially only appear at high-order rational flux-surfaces since the vacuum rotational transform avoids low-order rationals besides the plasma boundary where the rotational transform is close to $2/6$.}
	\label{fig:vac_flux}
\end{figure}

\begin{figure}[h]
	\centering
	\begin{subfigure}[t]{.32\textwidth}
		\centering
		\includegraphics[width=1.0\linewidth]{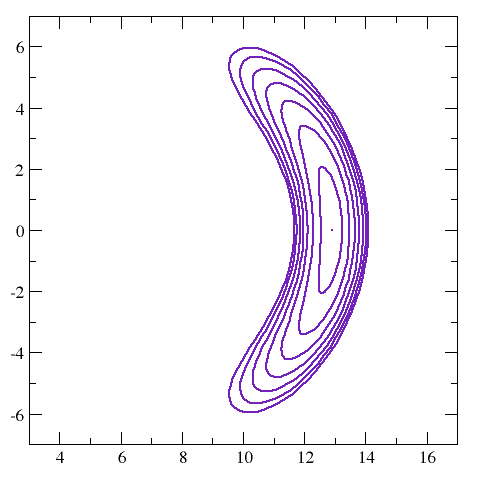}
		\caption{$\varphi =0$}
	\end{subfigure}
	\hfil
	\begin{subfigure}[t]{.32\textwidth}
		\centering
		\includegraphics[width=1.0\linewidth]{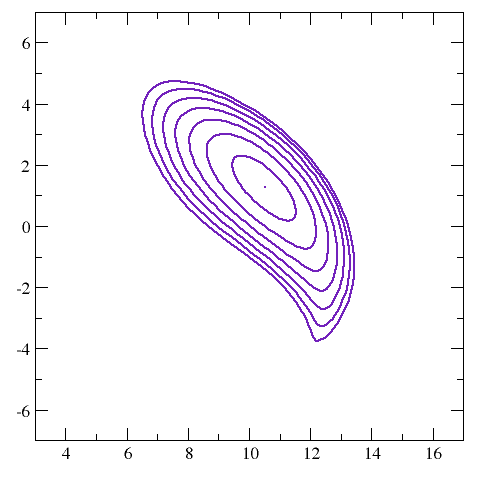}
		\caption{$\varphi = 45^\circ$}
	\end{subfigure}
	\hfil
	\begin{subfigure}[t]{.3325\textwidth}
		\centering
		\includegraphics[width=0.955\linewidth]{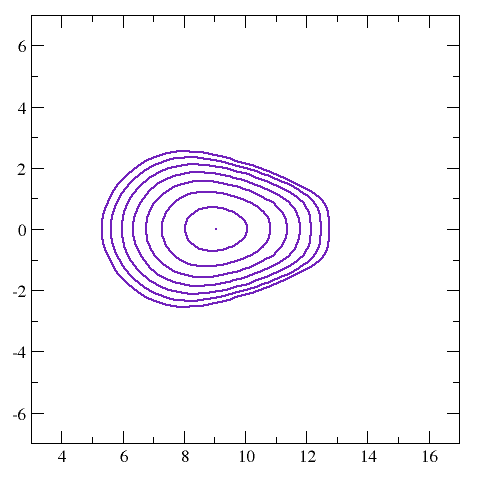}
		\caption{$\varphi= 90^\circ$}
	\end{subfigure}
	\caption{The flux surfaces at different toroidal angles $\varphi$ and with $\beta=3.5\%$ as calculated by VMEC.}
	\label{fig:poloidal_cuts}
\end{figure}

\begin{figure}[h]
	\centering
	\begin{subfigure}[t]{.4\textwidth}
		\centering
	\includegraphics[width=1.0\linewidth]{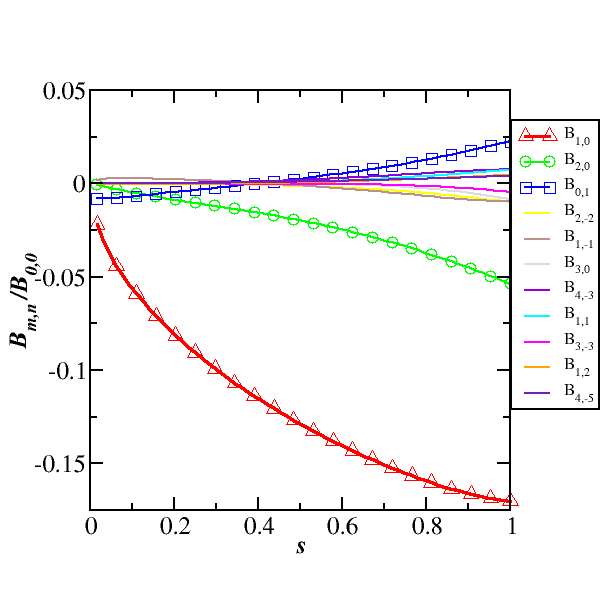}
	\caption{The Fourier spectrum of the magnetic field strength normalized to $B_{00}$.}
	\label{fig:bmnAll}
    \end{subfigure}
    \hfil
    \begin{subfigure}[t]{.4\textwidth}
    \centering
	\includegraphics[width=1.0\linewidth]{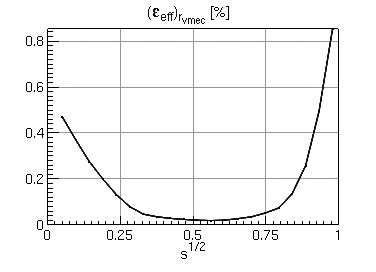}
	\caption{The effective ripple of the optimized QA equilibrium design.}
	\label{fig:eff_eps}
	\end{subfigure}
	\caption{Magnetic field strength spectrum left and the effective ripple profile right.}
\end{figure}

The configuration was optimized with an aspect ratio of 3.4, two field periods, a volume-averaged plasma beta of 3.5\%, and a pressure profile $\sim 1-0.8 s + 1.3 s^2 - 1.5 s^3$, see Fig.~\ref{fig:threeD}. The choice of two field periods facilitates the design of a modular coil set for low aspect ratios. The rotational transform profile of the optimized design lies between 0.3 and 0.5, see Fig.~\ref{fig:iota}, so that the low-order rational $\frac{1}{2}$ is avoided. The scaled plasma current has a total value of $2.5\,$MA at reactor size (V=$1900\,$m$^3$). The flat vacuum-rotational-transform profile is shown in Fig.~\ref{fig:vacIota}. It varies between 0.317 and 0.332 and therefore $8/25=0.32$ is the rational mode number with the smallest denominator in the vacuum-rotational-transform profile besides near the plasma boundary where it is close to $2/6\approx0.333$. The Poincar\'{e} plots of the vacuum magnetic field indicate nested magnetic surfaces, see Fig.~\ref{fig:vac_flux}. This is beneficial for the start-up of the device where there will be no plasma current to increase the rotational transform profile to its targeted profile. All vacuum parameters were calculated assuming a continuous current-carrying winding surface using the virtual-casing principle~\cite{VirtualCasing}. They therefore depend on the realization of discrete coils, which are not yet finalized. For $\beta=3.5\%$, the poloidal cross sections of the flux surfaces calculated with VMEC at the toroidal angle $\varphi =0,\, 45^\circ, \, 90^\circ$ are displayed in Fig.~\ref{fig:poloidal_cuts}. 
The magnetic field strength on the plasma boundary is shown in Fig.~\ref{fig:threeD}. Fig.~\ref{fig:contours} illustrates the (quasi-)axisymmetric magnetic field structure at half radius. The QA error is smallest around the flux surface $s=0.4$, but increases towards the plasma edge and towards the plasma axis, as one would expect from theory~\cite{Garren}. As mentioned before, the quasi-axisymmetric error was optimized on various flux surfaces between $s=0.25$ and $s=0.5$. This is also reflected in the Fourier spectrum of the magnetic field strength, see Fig.~\ref{fig:bmnAll} where all the components of the magnetic field strength normalized to $B_{00}$ are shown. All the non-quasi-axisymmetric components of the magnetic field strength are smaller than 2.5\% of $B_{00}$. The leading non-quasi-axisymmetric component is the mirror term $B_{01}$, which has a maximum near the plasma edge. The other components are smaller than 1\% throughout the entire plasma. The effective ripple $\epsilon_{eff}$, which is a measure of neoclassical transport~\cite{effeps}, attains values between 0.01\% at approximately half radius and 0.6\% at the edge, and lies below 0.2\% in most of the plasma volume, see Fig.~\ref{fig:eff_eps}. This is a side effect of the improved quasi-axisymmetry (since perfect quasi-symmetry implies $\epsilon_{eff}=0$) rather than the result of explicit optimization. 
\subsection{Fast-particle confinement}
\label{fast}
\begin{figure}[h]
	\centering
	\begin{subfigure}[t]{.4\textwidth}
		\centering
	\includegraphics[width=1.\linewidth]{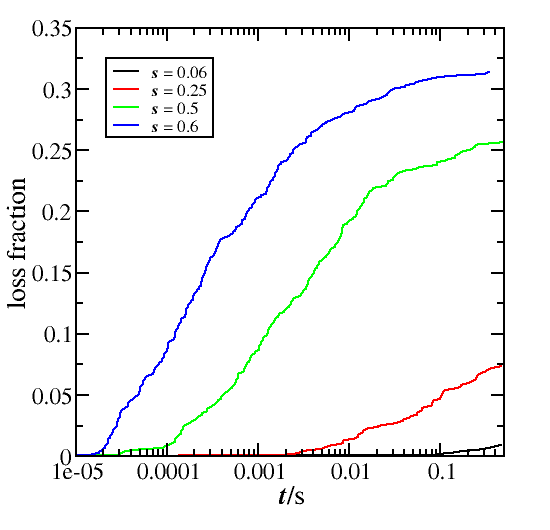}
	\caption{The fast particle loss fraction at flux surfaces between $s=0.06$ and $s=0.6$ with the original parameters and with a plasma volume of $1900$m$^3$. }
	\label{fig:loss_fraction}
	\end{subfigure}
	\hfil
	\begin{subfigure}[t]{.4\textwidth}
		\centering
		\includegraphics[width=1.4\linewidth]{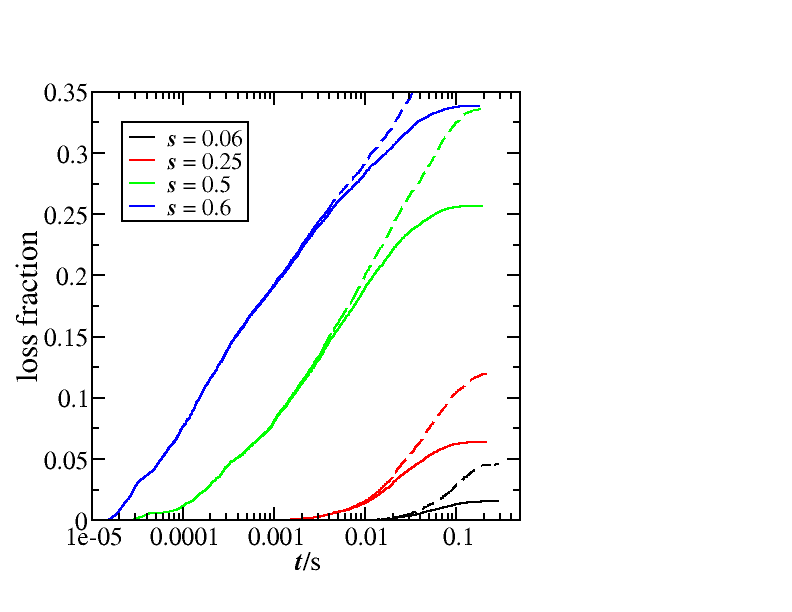}
		\caption{The loss fraction of fast particles (dashed lines) and energy (solid lines) including collisions with the background plasma and accounting for an electric field. }
		\label{fig:loss_fraction_Er}
	\end{subfigure}	
	\caption{Rates of fast-particle losses as calculated with ANTS.}
\end{figure}

To determine whether this magnetic configuration could be relevant for a reactor, the alpha-particle (with energies of $3.5\,$MeV) confinement was investigated. The fast-particle loss fraction was evaluated with the drift orbit code ANTS (plasmA simulatioN with drifT and collisionS)~\cite{Ants}. The configuration was scaled to reactor size with a volume of $1900\,$m$^3$, a major radius of $10.3\,$m, and a minor radius of $3.1\,$m, with a volume averaged magnetic field of $5\,$T. Two results are presented here: one where the guiding centre drifts are calculated without any collisions, Fig.~\ref{fig:loss_fraction}, and the other including collisions with the background plasma and an electric field, Fig.~\ref{fig:loss_fraction_Er}. In the case with collisions, specific density profiles of deuterium and tritium were imposed.  Using these profiles, the temperature and radial electric field were self-consistently determined following a similar procedure to that which is described in~\cite{NTSS2011}.
One thousand test particles, equally distributed on a flux surface, were launched with uniformly-distributed pitch angles and traced for half a second. 
In the collisionless case the loss fraction is below 1\% for the flux surface $s=0.06$ ($r\approx 0.25$) and for the flux surface $s=0.25$ (approximately half radius) it is 7.4\%, see Fig.~\ref{fig:loss_fraction}. 
In the case including collisions with the background plasma, the particles are lost more quickly than without collisions (because of collisional scattering onto collisionless loss orbits), however the energy losses are comparable to those without collisions, see Fig.~\ref{fig:loss_fraction_Er}. Note that the curves describing particle and energy losses start to diverge after approximately one slowing-down time, which is about $0.03\,$s for $s=0.06$.
\subsection{Stability}
\label{stab}
\begin{figure}[h]
	\begin{subfigure}[t]{.45\textwidth}
	\includegraphics[width=0.95\linewidth]{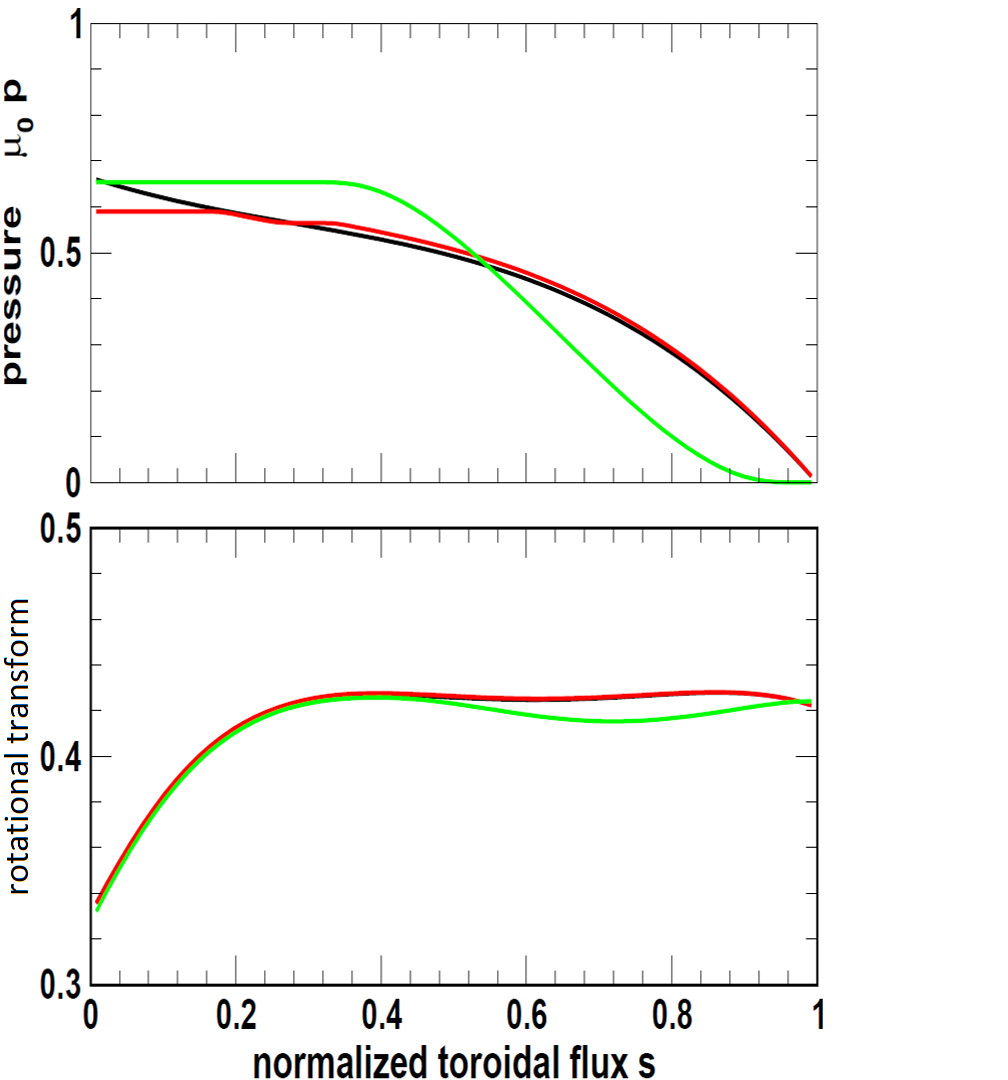}
	\caption{}
	\label{fig:pressure}
	\end{subfigure}
	\hfil
	\begin{subfigure}[t]{.45\textwidth}
	\includegraphics[width=1.0\linewidth]{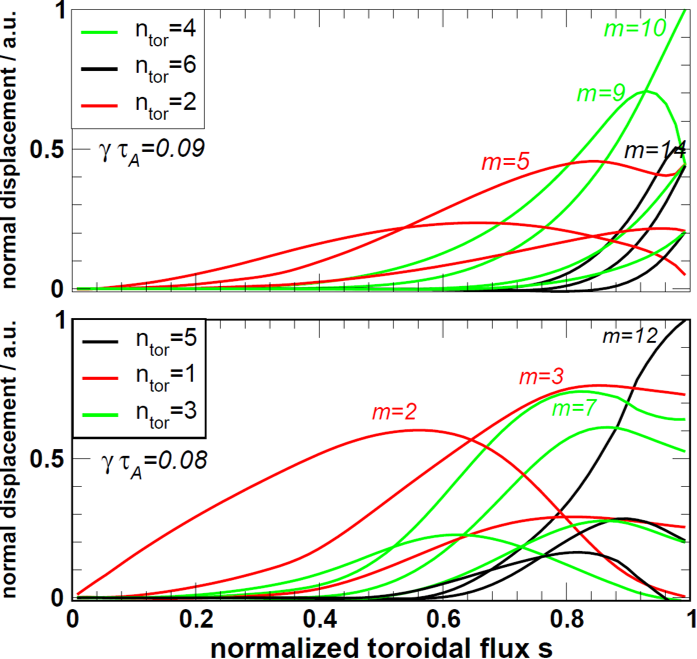}
	\caption{}
	\label{fig:unstable}
	\end{subfigure}
	\caption{Pressure profiles with related rotational transform profiles and modes structures for $\beta \approx 3.5\%$. (a) Top: black: un-flattened pressure profile; green, red: altered profiles. Bottom: associated rotational transform profiles with fixed current density profiles. (b) Unstable modes. Top: for the red pressure profile (flattened only near the magnetic axis). Bottom: mode structure for green pressure profile (flattened near magnetic axis and near edge), this mode structure is less local than that of the other case.}
\end{figure}

\begin{figure}[h]
	\centering
	\includegraphics[width=0.5\linewidth]{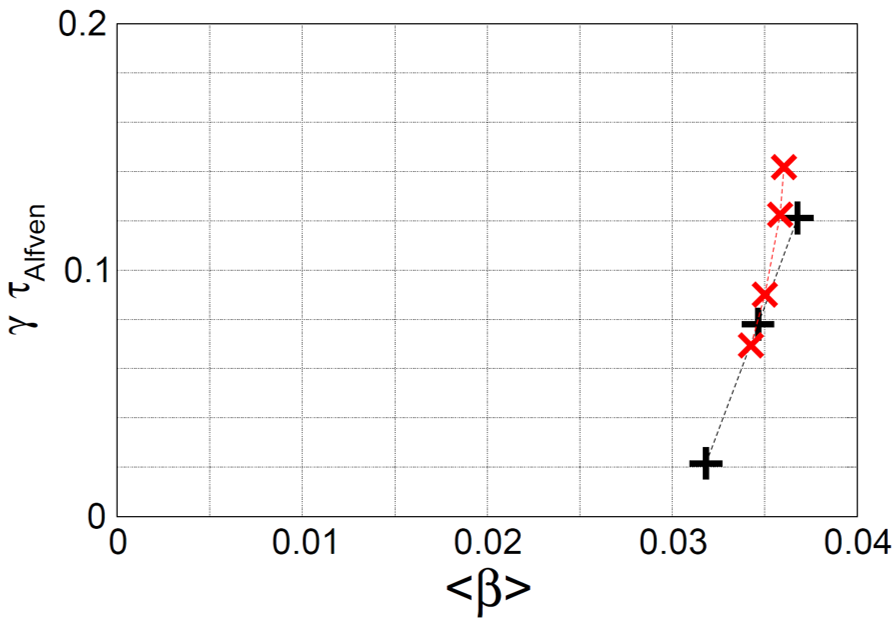}
	\caption{Ideal MHD stability $\beta$-limit: dependency of the growth rate on average plasma-beta. The red points correspond to the pressure profile which has been flattened near the magnetic axis. The black points correspond to the pressure profile flattened both at the magnetic axis and the plasma boundary (see Fig.~\ref{fig:pressure}). At $\langle\beta\rangle$ below 3\% ideal MHD stability prevails.}
	\label{fig:limit}
\end{figure}

As described above, one of the main advantages of QA configurations compared to tokamaks is their potential for being more stable and, hopefully, free from disruptions. The ideal MHD stability of the configuration was evaluated with the CAS3D code~\cite{CAS} in its free-boundary version. Three different pressure profiles were examined, see Fig.~\ref{fig:pressure}. For the stability calculation, the equilibrium pressure profiles were chosen to be flat, $p'(s)=0$, near the magnetic axis, where there are some low order rational surfaces, e.g. $\iota = 2/5$, in Fig.~\ref{fig:threeD} for the finite-$\beta$ case. In this way, diverging parallel current densities are eliminated that appear in ideal MHD equilibria instead of magnetic islands~\cite{Boozer81,Helander14}.
This flattening of the pressure profile only has a small effect on the rotational transform profile, see Fig.~\ref{fig:pressure}. 
 For $\beta \approx 3.5\%$ the plasma is ballooning unstable. The Fourier harmonics of the displacement in the direction normal to the flux surfaces are shown in Fig.~\ref{fig:unstable}. The ballooning nature of the modes, as well as their free-boundary nature, is evident from the high amplitudes near the plasma boundary.
By reducing the plasma beta and keeping the normalized profile fixed, a $\beta$-stability limit of 3\% is determined for the chosen numerical parameters, see Fig.~\ref{fig:limit}. This scan was performed with a fixed current profile and a fixed plasma boundary, Fig.~\ref{fig:poloidal_cuts}. 
\\
\subsection{Flux-surfaces with varying $\beta$ and current}
\label{spec}
Magnetic surfaces are not guaranteed to exist in general three-dimensional MHD equilibria without continuous symmetry~\cite{Rosenbluth66,Meiss1992}. While the vacuum field can be designed to possess nested magnetic surfaces (see Fig.~\ref{fig:vac_flux}), plasma currents are necessarily present at finite plasma pressure and thus the magnetic surfaces can potentially be destroyed at finite $\beta$. Moreover, the design presented herein considers a finite net toroidal current, $I_{\phi}$, and thus we must also assess whether this current can degrade (or improve) the quality of the confinement. Following the procedure described in Ref.~\cite{Loizu2017}, we use SPEC (Stepped-Pressure Equilibrium Code)~\cite{Hudson2012,SPEC_3D} in order to assess the possible formation of magnetic islands and magnetic-field-line chaos at different values of $\beta$ and $I_{\phi}$. A simplified pressure profile is assumed with $p(s)=p_0$ for $s\leq0.3$ and $p(s)=0$ for $s\geq0.3$. In this way, the pressure gradient and the pressure-driven (bootstrap) current density are concentrated on a single flux-surface. In each of the two volumes separated by that surface, SPEC allows the plasma to explore \emph{all possible} magnetic reconnection events that would lower the plasma potential energy. The equilibria thereby obtained with SPEC may not correctly describe the expected equilibria at given $\beta$ and $I_{\phi}$, but rather represent the ``worst-case-scenario" in which ``maximal relaxation" is allowed while supporting the prescribed pressure and current. We shall interpret the results as follows: whenever good flux surfaces are to be found despite the allowed relaxation, it shall be understood that no possible relaxation mechanism is likely to destroy these surfaces. When, on the other hand, islands and chaotic field-lines are produced, it shall be inferred that this is the Ôworst-case scenarioÕ and the potential destruction of flux-surfaces is subject to the available relaxation and healing mechanisms.

\begin{figure}[h]
	\centering
	\includegraphics[width=0.33\linewidth]{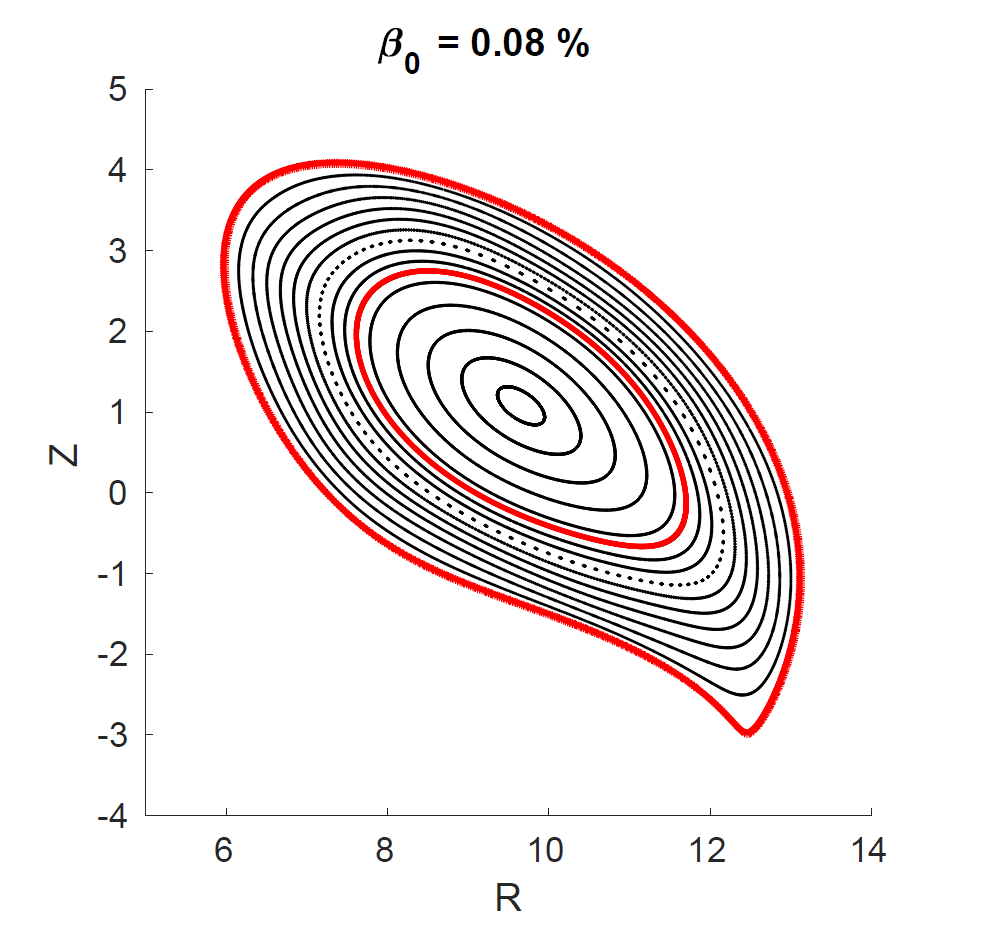}\hfill
	\includegraphics[width=0.33\linewidth]{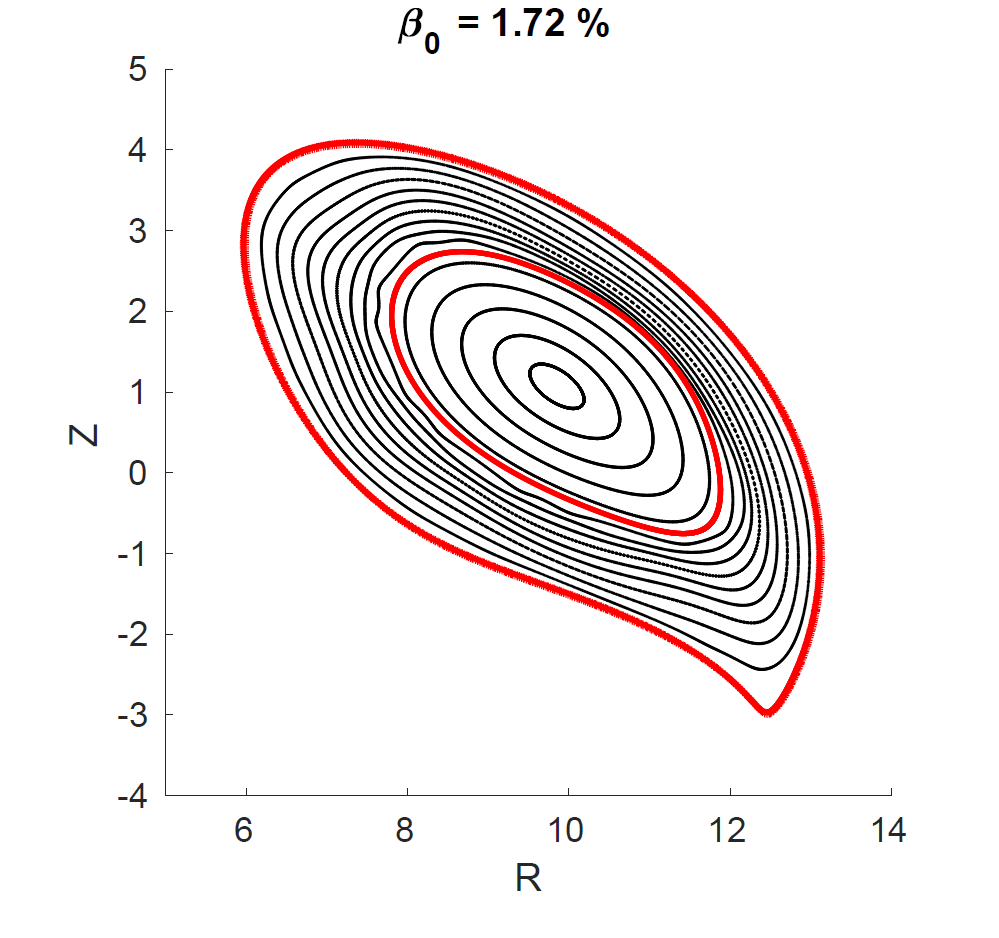}\hfill
	\includegraphics[width=0.33\linewidth]{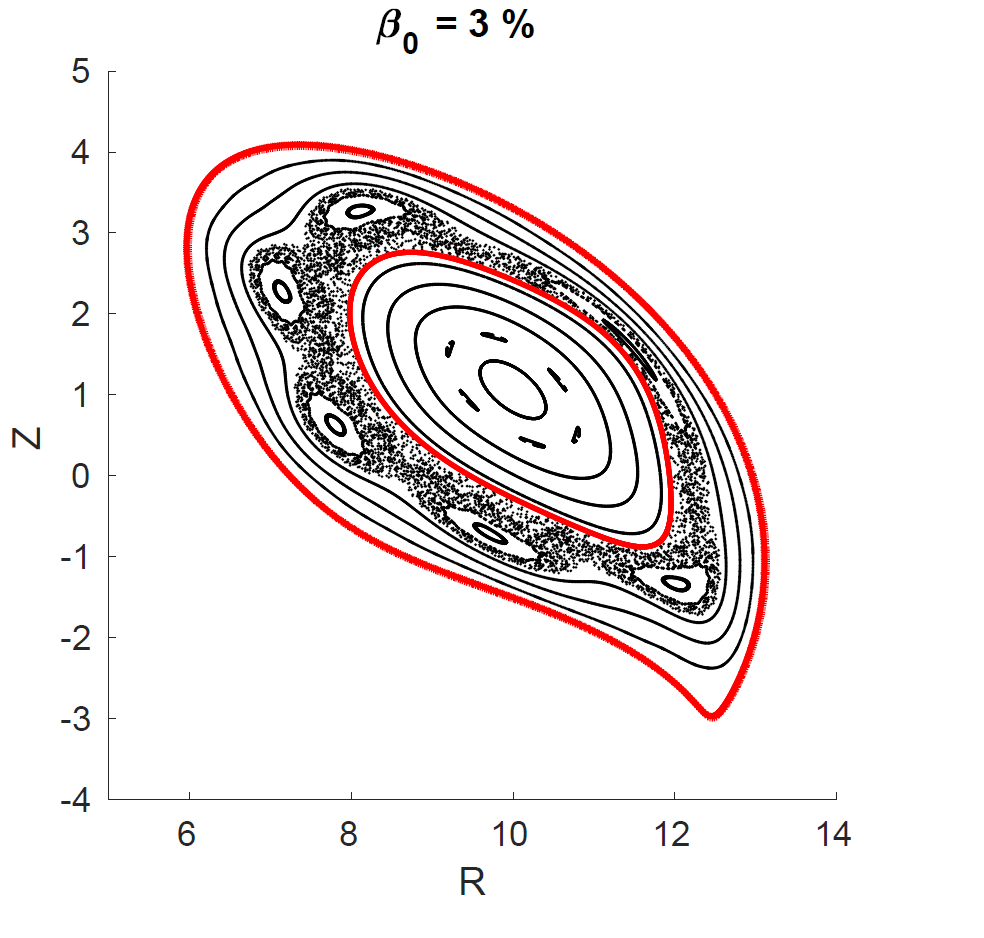}
	\caption{Poincar\'e plots of the poloidal cross-section at $\varphi=37.5^\circ$ and with $I_\phi=0$ at (from left to right) $\beta=0.08\%,\, 1.72\%$ and $3\%$ calculated by SPEC with the plasma boundary fixed.}
	\label{fig:beta}
\end{figure}

\begin{figure}[h]
	\centering
	\includegraphics[width=0.8\linewidth]{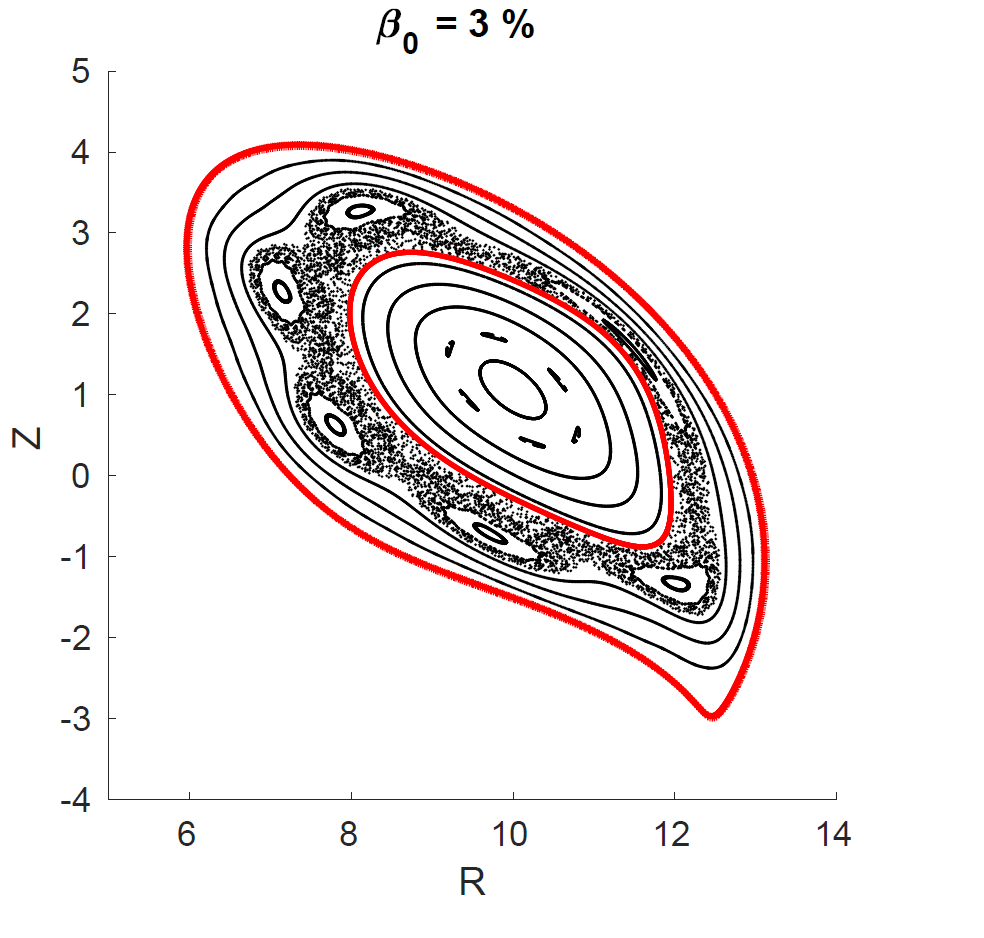}
	\caption{Enlarged Poincar\'e plot of the poloidal cross-section at $\varphi=37.5^\circ$ with $I_\phi=0$ at  $\beta=3\%$ calculated by SPEC with the plasma boundary fixed. A chain of six islands is visible.}
	\label{fig:betaislands}
\end{figure}

\begin{figure}[h]
	\centering
	\includegraphics[width=0.33\linewidth]{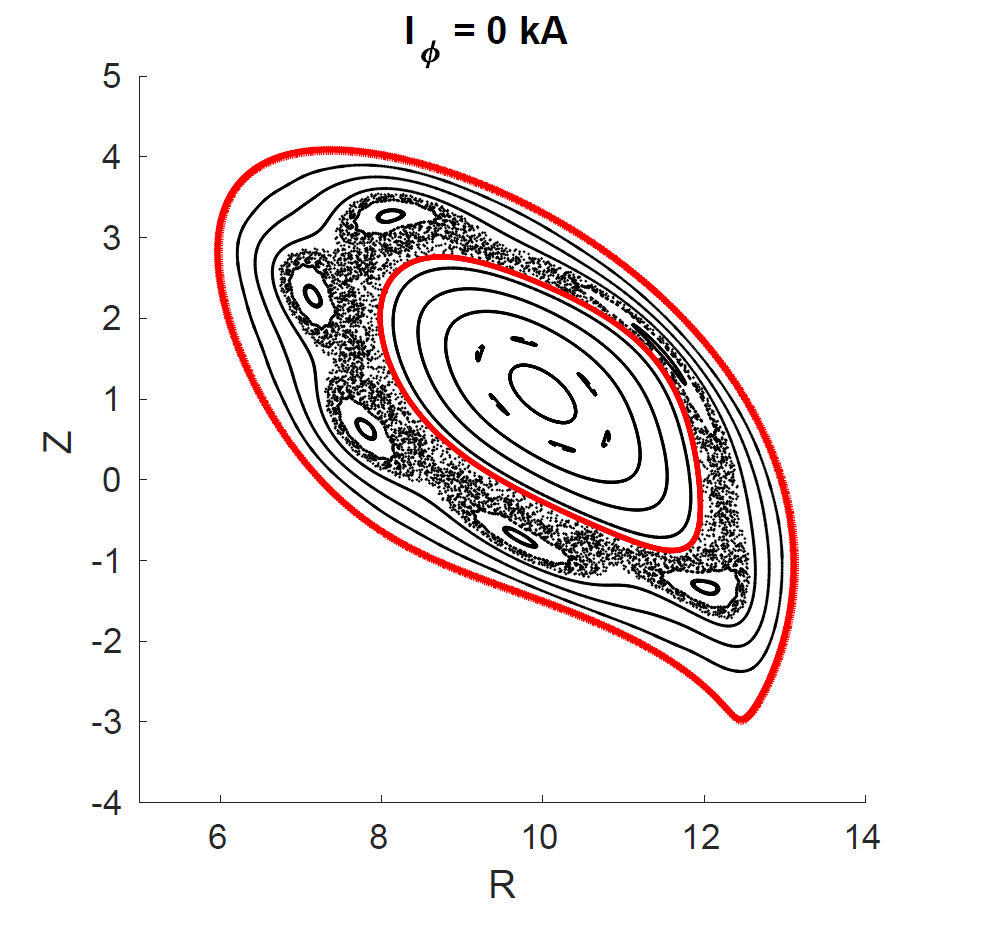}\hfill
	\includegraphics[width=0.33\linewidth]{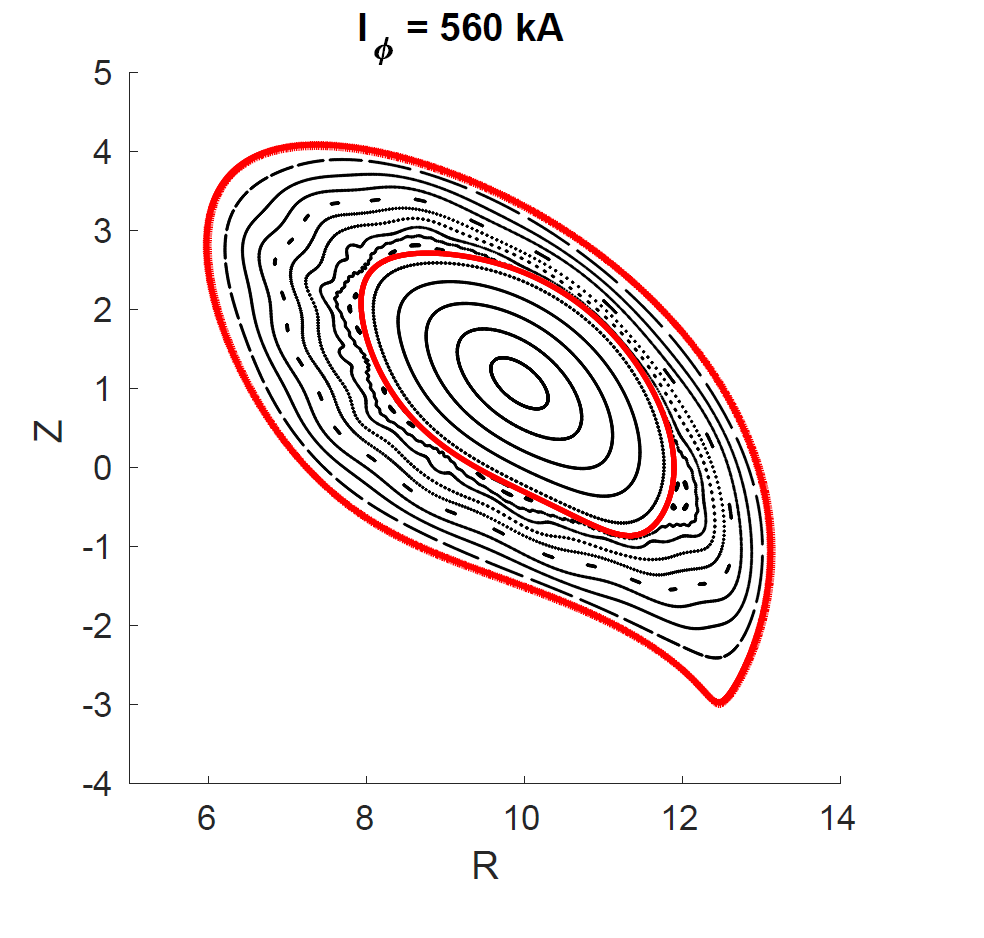}\hfill
	\includegraphics[width=0.33\linewidth]{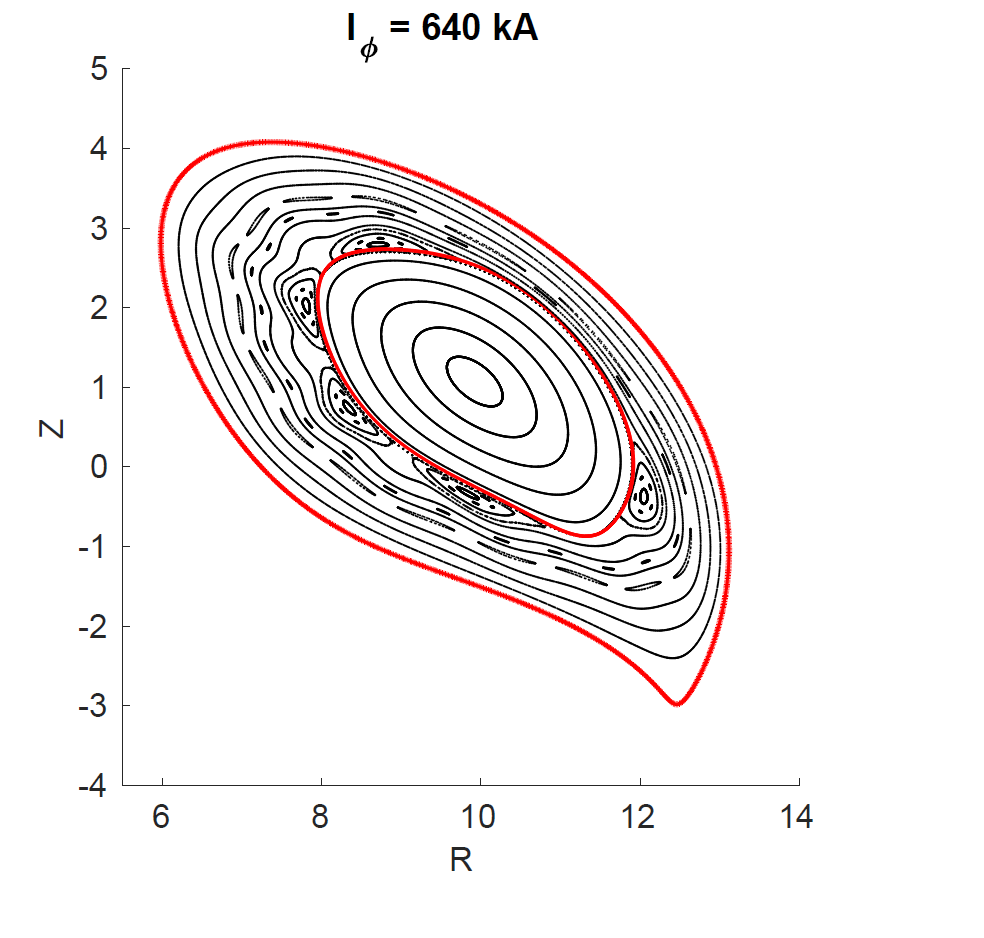}
	\caption{Poincar\'e plots of the poloidal cross-section at $\varphi=37.5^\circ$ and with constant $\beta=3\%$ and (from left to right) $I_\phi=0.0$kA, $560$kA and $640$kA calculated by SPEC with the plasma boundary fixed.}
	\label{fig:current}
\end{figure}

Fig.~\ref{fig:beta} shows examples of Poincar\'e sections of the equilibrium magnetic field obtained from SPEC calculations with increasing values of $\beta$, keeping $I_{\phi}=0$. An island chain emerges at around $\beta=1.5 \%$ and continuously increases in size. This is caused by the rotational transform crossing the resonance $\iota= 2/6=1/3$, see Fig.~\ref{fig:betaislands}. A second scan is performed in which the total enclosed net-toroidal-current $I_{\phi}$ is increased at fixed $\beta=3\%$. Some illustrative Poincar\'e sections for different values of $I_{\phi}$ are shown in Fig.~\ref{fig:current}. We remark that the effect of the net toroidal current on the rotational transform can be estimated as $\iota \sim \mu_0 I_{\phi} R / (2\pi a^2 B_{\phi})$, where $R$ is the major radius, $a$ is the minor radius, and $B_{\phi}$ is the toroidal magnetic field. Since the location of the current in SPEC calculations is at about half the minor radius, the rotational transform relevant for this design is obtained for $I_{\phi}\approx 600\,$kA at reactor size. Around this value of current, we observe that the original region of $\beta$-induced islands and chaos are reduced (hence the current improves the quality of confinement) although other resonances around $\iota=2/5$ appear soon thereafter. We conclude that the designed equilibrium may not have chaotic
regions but is clearly sensitive to the current within the plasma. Further investigations will be carried out in the future in order to gain confidence in these predictions. In particular, free-boundary SPEC calculations will be performed and the robustness of the magnetic topology will be assessed with respect to $\beta$ and $I_{\phi}$. \\
\subsection{Neoclassical transport and bootstrap current}
\label{neo}
\begin{figure}[h]
	\centering
	\begin{subfigure}[t]{.45\textwidth}
		\centering
		\includegraphics[width=1.0\linewidth]{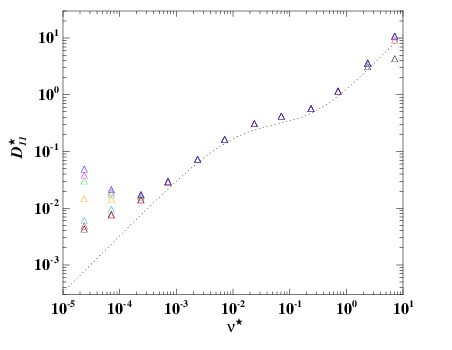} 
		\caption{Dependency of the mono-energetic radial transport coefficient $D*_{11}$ to collisionality. A banana regime is clearly visible at $\nu^*$ between $10^{-3}$ and $10^{-2}$.}
		\label{fig:d11}
	\end{subfigure}
	\hfil
	\begin{subfigure}[t]{.45\textwidth}
		\centering
		\includegraphics[width=1.0\linewidth]{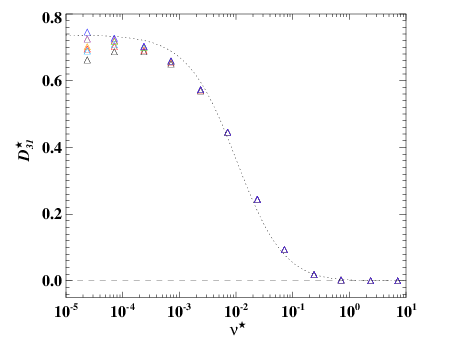} 
		\caption{The mono-energetic bootstrap current $D*_{31}$ coefficient versus collisionality. }
		\label{fig:d31}
	\end{subfigure}
	\caption{Mono-energetic transport coefficients~\cite{Craig} as calculated by DKES at $r=0.5$ with $|B_{m,n}>0.0001|$; triangular shapes: for the new equilibrium design for $E/vB$ between $3\times 10^{-6}$ and zero; dotted line: transport coefficient for an equivalent tokamak case.}
\end{figure}

The neoclassical mono-energetic transport coefficients~\cite{Craig} have been evaluated with DKES~\cite{DKES, DKES2}. Both the radial and bootstrap transport coefficients are close to the equivalent tokamak coefficients at half radius, see Figs.~\ref{fig:d11} and~\ref{fig:d31}. Only at very small collisionality $\nu^* < 10^{-3}$ do the coefficients depend on the electric field. A clear banana regime can be observed, which is relatively rare in stellarators~\cite{Craig} and provides further evidence for the minimization of the QA error near half-radius. At smaller and larger radii, the transport coefficients are more dependent on the electric field. This is in agreement with the effective ripple of this configuration, see Fig.~\ref{fig:eff_eps}. It is worth emphasizing that the reduction of the effective ripple is a by-product; it was not optimized explicitly but is a result of reducing the QA error between $s=0.25$ and $s=0.5$.\\
The bootstrap current has been evaluated with the NTSS code~\cite{NTSS2006} for a hydrogen plasma in a configuration scaled to the same volume as ASDEX-Upgrade of around $14.5\,$m$^3$ with a magnetic field of $2.5\,$T on axis and with a volume-averaged $\beta$ of 1.5\%. The obtained bootstrap current was around $226\,$kA which is close to the current used in the optimization (for this size it would be 250kA for $\beta\approx 3.5\%$). The rotational transform profile changes only slightly with the new bootstrap current. We can therefore conclude that the configuration behaves similarly to a tokamak and that the anticipated bootstrap current is of the right order. Finally, the next and final section presents an initial description of the coil design. 
\subsection{Coils}
\label{coils}
\begin{figure}[h]
	\centering
	\begin{subfigure}[t]{.45\textwidth}
		\centering
		\includegraphics[width=1.0\linewidth]{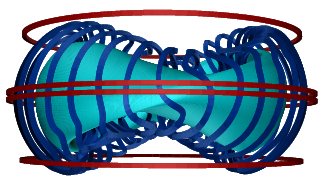} 
		\caption{Side view of preliminary coils for the new equilibrium design with the original plasma boundary.}
		\label{fig:coilsside}
	\end{subfigure}
	\hfil
	\begin{subfigure}[t]{.45\textwidth}
		\centering
		\includegraphics[width=1.\linewidth]{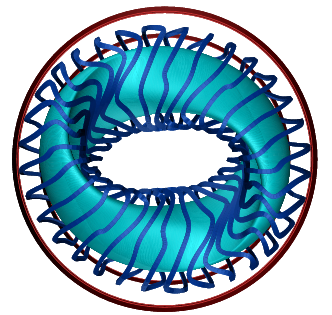} 
		\caption{Top view of preliminary coils for the new equilibrium design with the original plasma boundary.}
		\label{fig:coilstop}
	\end{subfigure}
	\caption{Preliminary set of coils: It is a set of 8 types of coils which leads to 32 modular coils with additional 4 poloidal field coils. The smallest radius of curvature appears near the quarter period poloidal cross section, due to the strong plasma edge shaping.}
	\label{fig:coils}
\end{figure}

A first investigation concerning practicability of modular coils for a reactor-sized configuration have been performed with the ONSET code~\cite{ONSET}. Four poloidal field coils have been employed with 8 types of modular coils corresponding to 32 modular coils in total, see Fig.~\ref{fig:coils}. The coils have a maximum relative magnetic field error (given by $e_l=|\mathbf{B}.\mathbf{n}|/|\mathbf{B}|$ with $\mathbf{B}$ the magnetic field on the plasma boundary and $\mathbf{n}$ the normal vector of the plasma boundary) of around 4.1\% and a mean relative magnetic field error ($\int \mathrm{d}A\,\, e_l/A$ with $A$ the surface of the plasma boundary) of 0.95\%. One can see that the most difficult shape of the coils is near the $\phi=45^\circ$ cross section, see Fig.~\ref{fig:cutmid}. Nevertheless, the clearance between the centreline of any two coils exceeds 51cm everywhere for a reactor-size machine and the minimum radius of curvature is $63\,$cm. This is an encouraging first step in the coil design process but further work has to be done. 
\section{Conclusion and future work}
A new quasi-axisymmetric stellarator configuration has been designed to possess a number of favourable features. It was found by optimizing the curvature of the plasma boundary, the magnetic well, rotational transform, and the quasi-axisymmetric field error. By varying the flux surface on which the quasi-axisymmetry was enforced, the optimization procedure was able to find configurations with particularly good neoclassical confinement.\\

The new plasma design has collisionless fast-particle loss rates below 8\% for flux surfaces $s \leq 0.25$. We examined the ideal-MHD stability with CAS3D and found that it has a stability limit of $\langle\beta\rangle\sim3\%$. The vacuum flux surfaces do not posses significant islands, but small ones appear when the current and beta are varied, but without the appearance of large stochastic regions. The neoclassical transport coefficients are nearly the same as in an equivalent tokamak at half radius. The weakest point of our optimization is probably the coil design, which is however still preliminary.\\

In future work, we aim to reduce the strong shaping at the plasma boundary at the mid-plane cross section to simplify the coil design. This might be possible by reducing the vacuum rotational transform while simultaneously checking MHD stability. Once improved coils have been found the confinement has to be verified.\\
Additionally, we will include a self-consistent bootstrap current into the optimization iteration to evaluate whether we can achieve all of these advantages.  Further optimization work could seek to include ideal-MHD stability calculations, and, if possible, check for stochastic fields inside the optimization loop. 
 An appropriate divertor concept will also need to be found, which would feed back into the plasma boundary optimization.\\
Further, it has to be investigated whether the relative large vacuum rotational transform is sufficiently large to avoid disruptions.
\\

\textbf{Acknowledgements}\\ This work has been carried out in the framework of the EUROfusion Consortium and has received funding from the Euratom research and training programme 2014-2018 under grant agreement no. 633053. The views and opinions expressed herein do not necessarily reflect those of the European Commission.

\bibliography{C:/Users/Sophia/Documents/PostDoc-Greifswald/Greifswald/MyPapers/quasi_axisymmetry_BP,C:/Users/Sophia/Documents/PostDoc-Greifswald/Greifswald/MyPapers/opt_coils}{}

\begin{thebibliography}{10}

\bibitem{SPEC_3D}
J.~Loizu, S.~R. Hudson, and C.~N{\"u}hrenberg.
\newblock Verification of the {SPEC} code in stellarator geometries.
\newblock {\em Physics of Plasmas}, 23(11):112505, 2016.

\bibitem{disruption1}
W~VII-A Team.
\newblock Stabilization of the (2, 1) tearing mode and of the current
  disruption in the {W VII–A} stellarator.
\newblock {\em Nuclear Fusion}, 20(9):1093, 1980.

\bibitem{W7AS}
M~Hirsch, J~Baldzuhn, C~Beidler, R~Brakel, R~Burhenn, A~Dinklage, H~Ehmler,
  M~Endler, V~Erckmann, Y~Feng, J~Geiger, L~Giannone, G~Grieger, P~Grigull, H-J
  Hartfuß, D~Hartmann, R~Jaenicke, R~König, H~P Laqua, H~Maaßberg,
  K~McCormick, F~Sardei, E~Speth, U~Stroth, F~Wagner, A~Weller, A~Werner,
  H~Wobig, S~Zoletnik, and for~the W7-AS~Team.
\newblock Major results from the stellarator {Wendelstein 7-AS}.
\newblock {\em Plasma Physics and Controlled Fusion}, 50(5):053001, 2008.

\bibitem{CTH1}
M.~D. Pandya, M.~C. ArchMiller, M.~R. Cianciosa, D.~A. Ennis, J.~D. Hanson,
  G.~J. Hartwell, J.~D. Hebert, J.~L. Herfindal, S.~F. Knowlton, X.~Ma,
  S.~Massidda, D.~A. Maurer, N.~A. Roberds, and P.~J. Traverso.
\newblock Low edge safety factor operation and passive disruption avoidance in
  current carrying plasmas by the addition of stellarator rotational transform.
\newblock {\em Physics of Plasmas}, 22(11):110702, 2015.

\bibitem{HrN88}
J.~N{\"u}hrenberg and R.~Zille.
\newblock Quasi-helically symmetric toroidal stellarators.
\newblock {\em Physics Letters A}, 129(2):113 -- 117, 1988.

\bibitem{HSX07}
J.~M. Canik, D.~T. Anderson, F.~S.~B. Anderson, K.~M. Likin, J.~N. Talmadge,
  and K.~Zhai.
\newblock Experimental demonstration of improved neoclassical transport with
  quasihelical symmetry.
\newblock {\em Phys. Rev. Lett.}, 98:085002, Feb 2007.

\bibitem{HSX07_PoP}
J.~M. Canik, D.~T. Anderson, F.~S.~B. Anderson, C.~Clark, K.~M. Likin, J.~N.
  Talmadge, and K.~Zhai.
\newblock Reduced particle and heat transport with quasisymmetry in the
  helically symmetric experiment.
\newblock {\em Physics of Plasmas}, 14(5):056107, 2007.

\bibitem{Boozer95}
A~H Boozer.
\newblock Quasi-helical symmetry in stellarators.
\newblock {\em Plasma Physics and Controlled Fusion}, 37(11A):A103, 1995.

\bibitem{Boozer81}
Allen~H. Boozer.
\newblock Plasma equilibrium with rational magnetic surfaces.
\newblock {\em The Physics of Fluids}, 24(11):1999--2003, 1981.

\bibitem{Helander14}
Per Helander.
\newblock Theory of plasma confinement in non-axisymmetric magnetic fields.
\newblock {\em Reports on Progress in Physics}, 77(8):087001, 2014.

\bibitem{HrN94}
J.~N{\"u}hrenberg, W.~Lotz, and S.~Gori.
\newblock Quasi-axisymmetric tokamaks.
\newblock {\em Theory of Fusion Plasmas Varenna 1994}, page~3, Editrice
  Compositori, Bologna, 1994.

\bibitem{Garabedian96}
P.~R. Garabedian.
\newblock Stellarators with the magnetic symmetry of a tokamak.
\newblock {\em Physics of Plasmas}, 3(7):2483--2485, 1996.

\bibitem{NCSX}
B.E. Nelson, L.A. Berry, A.B. Brooks, M.J. Cole, J.C. Chrzanowski, H.-M. Fan,
  P.J. Fogarty, P.L. Goranson, P.J. Heitzenroeder, S.P. Hirshman, G.H. Jones,
  J.F. Lyon, G.H. Neilson, W.T. Reiersen, D.J. Strickler, and D.E. Williamson.
\newblock Design of the national compact stellarator experiment ({NCSX}).
\newblock {\em Fusion Engineering and Design}, 66(Supplement C):169 -- 174,
  2003.
\newblock 22nd Symposium on Fusion Technology.

\bibitem{CHSqa}
S.~Okamura, K.~Matsuoka, S.~Nishimura, M.~Isobe, I.~Nomura, C.~Suzuki,
  A.~Shimizu, S.~Murakami, N.~Nakajima, M.~Yokoyama, A.~Fujisawa, K.~Ida,
  K.~Itoh, P.~Merkel, M.~Drevlak, R.~Zille, S.~Gori, and J.~N{\"u}hrenberg.
\newblock Physics and engineering design of the low aspect ratio
  quasi-axisymmetric stellarator {CHS}-qa.
\newblock {\em Nuclear Fusion}, 41(12), 2001.

\bibitem{ESTELL}
M.~Drevlak, F.~Brochard, P.~Helander, J.~Kisslinger, M.~Mikhailov,
  C.~N{\"u}hrenberg, J.~N{\"u}hrenberg, and Y.~Turkin.
\newblock {ESTELL}: A quasi-toroidally symmetric stellarator.
\newblock {\em Contributions to Plasma Physics}, 53(6):459--468, 2013.

\bibitem{ROSE}
M.~Drevlak, C.~B. Beidler, J.~Geiger, P.~Helander, and Y.~Turkin.
\newblock Optimisation of stellarator equilibria with {ROSE}.
\newblock {\em submitted to Nuclear Fusion}, 2018.

\bibitem{Brent}
R.~P. Brent.
\newblock {\em Algorithms for Minimization without Derivatives}, chapter~5.
\newblock Prentice-Hall, 1973.

\bibitem{VMEC}
S.~P. Hirshman and J.~C. Whitson.
\newblock Steepest descent moment method for three dimensional
  magnetohydrodynamic equilibria.
\newblock {\em The Physics of Fluids}, 26(12):3553--3568, 1983.

\bibitem{NESCOIL}
P.~Merkel.
\newblock Solution of stellarator boundary value problems with external
  currents.
\newblock {\em Nuclear Fusion}, 27(5):867, 1987.

\bibitem{VM2MAG}
Matthias Borchardt.
\newblock private communication.

\bibitem{NTSS2006}
Yu. Turkin, H.~Maassberg, C.~D. Beidler, J.~Geiger, and N.~B. Marushchenko.
\newblock Current control by {ECCD} for {W7-X}.
\newblock {\em Fusion Science and Technology}, 50(3):387--394, 2006.

\bibitem{Garren}
D.~A. Garren and A.~H. Boozer.
\newblock Existence of quasihelically symmetric stellarators.
\newblock {\em Physics of Fluids B: Plasma Physics}, 3(10):2822--2834, 1991.

\bibitem{Plunk2018}
G.~G. Plunk and Per Helander.
\newblock Quasi-axisymmetric magnetic fields: weakly non-axisymmetric case in a
  vacuum.
\newblock {\em Journal of Plasma Physics}, 84(2):905840205, 2018.

\bibitem{VirtualCasing}
V.D. Shafranov and L.E. Zakharov.
\newblock Use of the virtual-casing principle in calculating the containing
  magnetic field in toroidal plasma systems.
\newblock {\em Nuclear Fusion}, 12(5):599, 1972.

\bibitem{effeps}
V.~V. Nemov, S.~V. Kasilov, W.~Kernbichler, and M.~F. Heyn.
\newblock Evaluation of 1/$\nu$ neoclassical transport in stellarators.
\newblock {\em Physics of Plasmas}, 6(12):4622--4632, 1999.

\bibitem{Ants}
M.~Drevlak, J.~Geiger, P.~Helander, and Y.~Turkin.
\newblock Fast particle confinement with optimized coil currents in the {W7-X}
  stellarator.
\newblock {\em Nuclear Fusion}, 54(7), 2014.

\bibitem{NTSS2011}
Y.~Turkin, C.~D. Beidler, H.~Maaßberg, S.~Murakami, V.~Tribaldos, and
  A.~Wakasa.
\newblock Neoclassical transport simulations for stellarators.
\newblock {\em Physics of Plasmas}, 18(2):022505, 2011.

\bibitem{CAS}
Carolin Schwab.
\newblock Ideal magnetohydrodynamics: Global mode analysis of three dimensional
  plasma configurations.
\newblock {\em Physics of Fluids B: Plasma Physics}, 5(9):3195--3206, 1993.

\bibitem{Rosenbluth66}
M.N. Rosenbluth, R.Z. Sagdeev, J.B. Taylor, and G.M. Zaslavski.
\newblock Destruction of magnetic surfaces by magnetic field irregularities.
\newblock {\em Nuclear Fusion}, 6(4):297, 1966.

\bibitem{Meiss1992}
J.~D. Meiss.
\newblock Symplectic maps, variational principles, and transport.
\newblock {\em Rev. Mod. Phys.}, 64:795--848, Jul 1992.

\bibitem{Loizu2017}
J.~Loizu, S.~R. Hudson, C.~N{\"u}hrenberg, J.~Geiger, and P.~Helander.
\newblock Equilibrium $\beta$-limits in classical stellarators.
\newblock {\em Journal of Plasma Physics}, 83(6):715830601, 2017.

\bibitem{Hudson2012}
S.~R. Hudson, R.~L. Dewar, G.~Dennis, M.~J. Hole, M.~McGann, G.~von Nessi, and
  S.~Lazerson.
\newblock Computation of multi-region relaxed magnetohydrodynamic equilibria.
\newblock {\em Physics of Plasmas}, 19(11):112502, 2012.

\bibitem{Craig}
C.D. Beidler, K.~Allmaier, M.Yu. Isaev, S.V. Kasilov, W.~Kernbichler, G.O.
  Leitold, H.~Maaßberg, D.R. Mikkelsen, S.~Murakami, M.~Schmidt, D.A. Spong,
  V.~Tribaldos, and A.~Wakasa.
\newblock Benchmarking of the mono-energetic transport coefficients - results
  from the international collaboration on neoclassical transport in
  stellarators ({ICNTS}).
\newblock {\em Nuclear Fusion}, 51(7):076001, 2011.

\bibitem{DKES}
S.~P. Hirshman, K.~C. Shaing, W.~I. van Rij, C.~O.~Beasley Jr., and E.~C.~Crume
  Jr.
\newblock Plasma transport coefficients for nonsymmetric toroidal confinement
  systems.
\newblock {\em The Physics of Fluids}, 29(9):2951--2959, 1986.

\bibitem{DKES2}
W.~I. van Rij and S.~P. Hirshman.
\newblock Variational bounds for transport coefficients in three‐dimensional
  toroidal plasmas.
\newblock {\em Physics of Fluids B: Plasma Physics}, 1(3):563--569, 1989.

\bibitem{ONSET}
Michael Drevlak.
\newblock Automated optimization of stellarator coils.
\newblock {\em Fusion Technology}, 33(2):106--117, 1998.

\end{thebibliography}
\bibliographystyle{unsrt}
\end{document}